# Earth and Space Science

**Key Points:**
- Multi-modal probabilistic inversion of geophysical logs using a single evaluation of a deep neural network
- Deep neural network outputs likely stratigraphic inversions and predictions ahead of data and their probabilities
- The model predicts more accurate and realistic solutions compared to the single-mode predictor

**Correspondence to:**
S. Alyaev,
saly@norceresearch.no

**Author Contributions:**
**Conceptualization:** Sergey Alyaev, Ahmed H. Elsheikh
**Data curation:** Sergey Alyaev
**Formal analysis:** Sergey Alyaev, Ahmed H. Elsheikh
**Methodology:** Sergey Alyaev, Ahmed H. Elsheikh
**Software:** Sergey Alyaev
**Validation:** Sergey Alyaev
**Visualization:** Sergey Alyaev
**Writing – original draft:** Sergey Alyaev
**Writing – review & editing:** Sergey Alyaev, Ahmed H. Elsheikh

## Direct Multi-Modal Inversion of Geophysical Logs Using Deep Learning

Sergey Alyaev[1]  and Ahmed H. Elsheikh[2]

[1]NORCE Norwegian Research Centre, Bergen, Norway, [2]Heriot-Watt University, Edinburgh, UK

**Abstract** Geosteering of wells requires fast interpretation of geophysical logs which is a non-unique inverse problem. Current work presents a proof-of-concept approach to multi-modal probabilistic inversion of logs using a single evaluation of an artificial deep neural network (DNN). A mixture density DNN (MDN) is trained using the "multiple-trajectory-prediction" loss functions, which avoids mode collapse typical for traditional MDNs, and allows multi-modal prediction ahead of data. The proposed approach is verified on the real-time stratigraphic inversion of gamma-ray logs. The multi-modal predictor outputs several likely inverse solutions/predictions, providing more accurate and realistic solutions compared to a deterministic regression using a DNN. For these likely stratigraphic curves, the model simultaneously predicts their probabilities, which are implicitly learned from the training geological data. The stratigraphy predictions and their probabilities obtained in milliseconds from the MDN can enable better real-time decisions under geological uncertainties.

**Plain Language Summary** Positioning the wells relative to geological targets and adjusting trajectory in real-time requires fast interpretation of streamed geophysical measurements. As such interpretations are not unique, high-quality decision-making requires the exploration of all likely interpretations and estimation of their probabilities. This study presents a mixture density deep neural network that correlates the log of the drilled well with the offset well and outputs a chosen number of interpretations of the geometry of geological layers and their probabilities. Moreover, by learning the likely configurations in the training geological data, one can extrapolate the interpretations ahead of the data. The presented model achieves good prediction accuracy while producing more realistic interpretations compared to the deterministic single-output model.

## 1. Introduction

Most of the inverse problems related to the interpretation of geophysical measurements are ill-posed. While in some problems, it is possible to use regularization to find a single physically viable solution, for many keeping several likely solution modes and estimating their probabilities is useful.

Fast probabilistic inversion of geophysical data is specifically crucial for processing measurements during drilling since it opens the possibility to use probabilistic decision algorithms (Alyaev et al., 2019; Chen et al., 2015; Kullawan et al., 2018) to adjust the trajectory during operation, resulting in better well placement. The intentional adjustment of the drilling trajectory termed geosteering, is essential for hydrocarbon wells (Bristow, 2000), but is gradually introduced in other types of drilling, for example, geothermal (Ungemach et al., 2021) and civil wells and tunnels (Johnson et al., 2021).

The most widespread method is stratigraphic-based geosteering inversion, which assumes that the unknown geomodel can be represented as curved and possibly faulted geological layers (strata), and the geophysical log response only depends on the vertical position within the stratigraphy. For the inversion, the geophysical logs from a geosteered horizontal well and corresponding logs from an offset vertical well are projected onto the stratigraphic geomodel. The depth and the lateral shape of the stratigraphic geomodel are then modified until a match of logs is obtained (Tadjer et al., 2021). In most operations, the matching is done visually by an expert, and only a single inversion is considered.

Recently, several authors developed methods for probabilistic real-time stratigraphic inversion. Winkler (2017) proposed depth correlation using Bayesian networks. Solving the resulting network with simplifying assumptions takes several minutes, but the algorithm can produce multi-modal distributions of likely solutions. Arbus and Wilson (2019) proposed a search-based inversion method with a database structured as a graph containing signatures of different stratigraphic layers. The look-up takes several minutes, and cloud solutions are necessary for real-time performance. Gee et al. (2020) and Maus et al. (2020) developed a method for tracking several solutions using stratigraphic misfit heatmaps. Their multi-modal inversion is proprietary and takes about a minute. Veettil and Clark (2020) adapted a sequential Monte Carlo/particle filter for solving the stratigraphic inversion recursively. The particle filters can approximate an arbitrary probability density function but rely on a heuristic model of stratigraphy changes. The quality of the prediction depends on the number of simulated particles, which in turn influences the computational time.

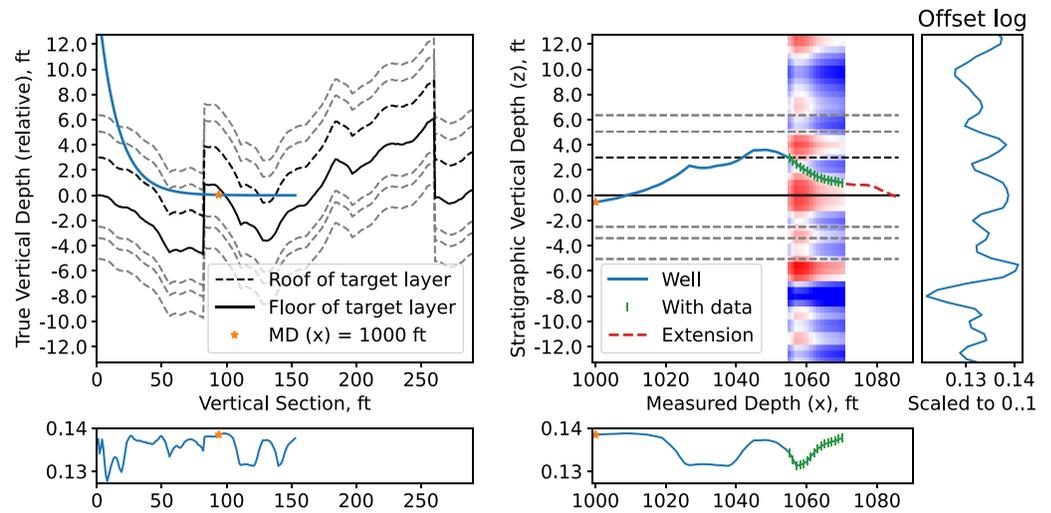

**Figure 1.** Left part: A realization of stratigraphy from the test dataset (Alyaev, 2022) with an example of a log for inclined well in vertical section/true vertical depth (VS/TVD) coordinates. Right part: Model problem. Part of the well from the left panel in MD/SVD coordinates is linked to stratigraphy and hence the offset-well log. The lower panels show the log corresponding to the displayed well, while the right-most panel shows the offset well log assumed to be invariant to the SVD coordinates. The heat-map image on the left is produced by subtracting the offset log from the well-log with zeros shown as white curves.

DNNs present a faster alternative to classical inversion techniques (Yu & Ma, 2021). MDNs directly output a probability distribution defined as a sum of learned kernels (inversion modes) such as Gaussians multiplied by their probabilities (Earp & Curtis, 2020). Unlike particle filters, MDNs do not require explicit heuristic proposal distributions but instead learn the most likely configurations from the training data. Previous attempts to use MDNs for Bayesian inversion in geophysics suffered from unstable training, resulting in essentially a single-mode prediction (Earp & Curtis, 2020; Earp et al., 2020; Meier et al., 2007). Therefore, Zhang and Curtis (2021) recommended an alternative approach based on invertible neural networks, which come with a computational overhead.

Here, we propose to use a modified MDN. Our supervised training uses the more stable "Multiple-Trajectory-Prediction" (MTP) loss function, which is, to our knowledge, the first adaptation of such techniques within the geosciences. Our MDN takes the vertical and the horizontal well-logs as input and outputs a chosen number of solutions and their probability based on a single evaluation done within milliseconds. Moreover, the training allows extending the stratigraphic inversion into a multi-modal probabilistic prediction ahead of measurements, enabling even easier integration with decision optimization methods.

## 2. Model Problem

During geosteering, the typical problem is finding the well's location relative to geological objects based on real-time data. When the subsurface can be represented as a layer-cake model with constant thickness the inverse problem is to find the well trajectory function in the depth coordinate relative to the stratigraphic layers, termed stratigraphic vertical depth (SVD). Given that we know the well trajectory in absolute coordinates: horizontal vertical section (VS), and vertical true vertical depth (TVD), as a function as measured depth (MD) along the well; finding the SVD well trajectory function $b^*$ (SVD function for short) is sufficient to reconstruct the geological model, see Figure 1.

Let $f(z)$ be a shallow geophysical well-log obtained from a vertical offset well, where $z$ is the SVD coordinate. Let $g(x)$ be the same kind of well-log obtained from a horizontal well, where $x$ is the MD equivalent to VS for the horizontal part. The stratigraphic inversion can be formalized as follows: Find SVD function $z_0 \leq b^*(x) \leq z_n$, such that:

$$f(b^*(x)) = g(x) \quad x \in [x_0, x_l], \tag{1}$$

where $z_0$ and $z_n$ are the SVD limits; $x_0$ and $x_l$ define the interval with measurements in MD. The SVD function $b^*(x)$ corresponds to a zero iso-line in $g(x)/f(z)$ heat-map image (which is similar to heat-map images defined in Maus et al. [2020]), see Figure 1.

When the data is limited to a few or only one shallow log (as in this case), there are often many zero iso-lines, several of which can be plausible solution modes. In view of this multi-modality, it is desirable to find several likely solutions $b^m(x)$ and their corresponding probabilities $p^m$. The probabilities should be estimated based on how common is such solutions in the chosen geological setting. To achieve proactive geosteering, the solutions should be considered together with predictions ahead of data, that is, find $b^m(x)$ for $x \in [x_0, x_{l_+}]$, where $x_{l_+} > x_l$. The prediction part would have much higher uncertainty than the inversion part of the SVD function.

## 3. DNN Architecture and MTP Loss

This section describes the proposed solution to the inverse problem using supervised learning of MDNs with a particular MTP loss function.

### 3.1. Inputs

Our model utilizes both well-logs $f(z)$ and $g(x)$ as inputs. This work limits the architecture to the fixed discrete interval with a regular grid for both $z$ and $x$. We define a reference vertical data vector as

$$f_k = f(z_k), \quad z = z_0, \ldots, z_n, \tag{2}$$

and assume that the horizontal well-log vector $g_j$ can be approximated as a function of $f_k$ using linear interpolation:

$$g_j = g(x_j) = f(b^*(x_j)) := (b_j - z_k) f_{k+1} + (z_{k+1} - b_j) f_k + \epsilon, \quad x = x_0, \ldots, x_l, \tag{3}$$

where $b_j = b^*(x_j)$ lies between $z_k$ and $z_{k+1}$; and $\epsilon$ is possible measurement noise. Note that in most of this initial study, the noise is set to zero during the training, validation, and testing of the reference model. However, we study the impact of noise during evaluation in Section 5.4 and training in Section 5.5.

Before sending the data to a DNN, the input vectors are pair-wise subtracted from each other forming the heat-map image (see Figure 1) with 'pixel' values $r_{i,j}$:

$$r_{k,j} = R_{k,j}(f, g) = f_k - g_j. \tag{4}$$

### 3.2. Outputs

Expanding on the ideas used to predict behavior in traffic (Cui et al., 2019), we adopt the output format. Our DNN predicts a predefined number of realizations of the SVD functions $b^m = \left[b_m, b_1^m, \ldots, b_{l_+}^m\right]^T$, as well as their (unscaled) logarithmic probabilities $\rho^m$. From a pragmatic engineering point of view, the true answer should closely match one of the realizations with the predicted probability. From a rigorous statistical perspective, the output of our MDN represents the mean curves of the selected (in our case, non-Gaussian) kernels with non-local support. The kernel is dictated by the regression part of the loss function, see Section 3.4. Unlike the classical MDN, we do not include an explicit kernel width/standard deviation model in the current work. Currently, it is fixed by the preset weighting factor of the loss contributions. In our implementation, the output of the DNN is a concatenated vector $P$ of $M$ predicted paths of length $l_+$ and the probability of each path:

$$P = \left[b_0^0, b_1^0, \ldots, b_{l_+}^0, \rho^0, \ldots, b_0^m, b_1^m, \ldots, b_{l_+}^m, \rho^m\right]^T, \tag{5}$$

where the upper indices correspond to the number of the mode, and the lower indices for $b^i$ correspond to the position along the predicted path. Thus the total size of outputs is $M * (l_+ + 1)$.

### 3.3. Model Architecture

The architecture of the model consists of two parts: image pre-processing Regression Head (RH) and multi-modal predictor (MDN), which can be written as an equation:

$$P = MDN \circ RH \circ R(f, g), \tag{6}$$

where $\circ$ is the function composition operator.

First, we extract information from images with an RH. To improve training, the input images' pixels are normalized with constants so that the dataset pixels will be distributed as the standard normal distribution. This DNN contains three pairs of 2D convolution ($3 \times 3$ filter) and max-pooling ($2 \times 2$) layers to process and downscale the image. They are followed by two fully connected layers, which produce features for the MDN. All layers in the RH use ReLU activation.

Second, a fully-connected DNN forms an MDN predictor. We adopt the dense architecture from Cui et al. (2019) which contains a sequence of two fully connected layers: a 4096-neuron-wide layer with ReLU activation and a linear output layer, which scales all $M * (m_+ + 1)$ outputs. In our current implementation, the division of the DNN into the RH and the MDN is arbitrary. In a general case, extra contextual information can be passed to the predictor as suggested by Cui et al. (2019). The current architecture gave sufficiently good results for this pre-study, but further hyper-parameter optimization to improve the model performance remains a part of future work.

### 3.4. Loss Function

To enable probabilistic multi-modal prediction of well-logs, we adopt an MTP loss function introduced by Cui et al. (2019). The MTP loss function $I_{MTP}$ consists of two parts: classification loss $I_{class}$ which accounts for the probability of modes, and best-mode regression loss $I_{reg}$, which accounts for the prediction quality:

$$I_{MTP} = \alpha_{class} I_{class} + I_{reg}. \tag{7}$$

The constant $\alpha_{class}$ balances the influence of loss function components. It can also be interpreted as the standard deviation within the predicted modes, that is, the width of a mode. High values $\alpha_{class}$ make modes wider, which may cause mode collapse, while low values might make modes too narrow, thus requiring many more modes to resolve an uncertain distribution. As such, the value of hyperparameter $\alpha_{class}$ has to be investigated for each problem.

For a sample $(r*, b*)$, the loss function tries to propagate only the best matching mode by first finding the mode $m*$ that is close to the reference data $b*$:

$$m^* = \underset{1 \le m \le m_+}{\mathrm{argmin}} \|b^* - b^m\|, \tag{8}$$

where $\| * \|$ is a chosen norm. In this study, we use 1-norm everywhere, which results in mean-absolute-error (MAE) distance. The contributions to the MTP loss functions modify the weights that influence the found best mode:

The probability of the mode $m*$ is maximized using a log-softmax function typical for classification:

$$I_{class} = -\log\left(\frac{\exp(\rho_{m^*})}{\sum_m \exp(\rho_m)}\right). \tag{9}$$

Note that this corresponds to the calibration of the other modes' probabilities given the summation in the denominator.

The prediction or regression loss reduces the MAE distance from the mode $m*$ to the true solution:

$$I_{reg} = \frac{\|b^* - b^{m^*}\|}{l^+}. \tag{10}$$

## Offset gamma-ray log

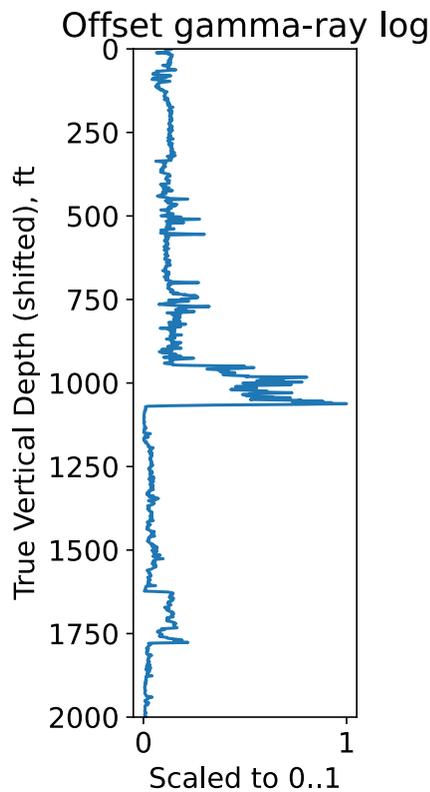

**Figure 2.** The full offset gamma-ray well-log adopted from Miner et al. (2021).

Our implementation uses the vector length normalization $l^+$ to simplify the comparison between experiments.

Rupprecht et al. (2017) prove that this type of loss function results in a mathematically sound approximation of probability density expressed as a mixture of kernels for the training of MDNs.

## 4. Dataset and Training

For the DNN training, we use a supervised learning process with a fixed dataset. This section describes the synthetic dataset used in this study and then the training implementation and convergence.

### 4.1. Synthetic Dataset

The synthetic dataset used for this study consists of a large set of stratigraphy realizations and an offset well-log file.

The stratigraphy-realization dataset consists of randomly generated SVD functions $b*(x)$ which follow a known trend (here zero) (Alyaev, 2022), see, for example, Figure 1. The offset well-log used for this study is the gamma-ray log from the Geosteering World Cup 2020 semi-finals, the unconventional well (Miner et al., 2021; Tadjer et al., 2021). This synthetic well-log is built based on observation in the Middle Woodford formation, located in the South-Central Oklahoma Oil Province in the United States. The well-log is discretized every 0.5 ft, and we normalize the values of the well-log to 0–1 interval, see Figure 2.

The training data consists of triples: a reference offset-well log, which is trimmed randomly to a short section of 64 cells (32 ft TVD); a sample of $b*(x)$ with 32 points (32 ft); and an observed well-log corresponding to the first 16 ft of $b*(x)$, obtained using formula 3. Every second sample of $b*(x)$ is centered around its start $b_c^*(x) = b*(x) - b*(x_0)$. Such centering gives more weight (in probability) to the inversions, for which we know the starting point from the previous inversion.

The complete training dataset contains 28 million samples. Additionally, we use two different datasets generated with the same rules: the validation dataset of 12 thousand samples and the testing dataset of 10,000 samples.

### 4.2. Training Setup

For the training, we use the loss function with $\alpha_{class} = 0.1$. The training uses the Adam optimizer (Kingma & Ba, 2014) from PyTorch (Paszke et al., 2019) with a learning rate set to $0.5e - 4$. The dataset is split into batches, each containing 512 samples. To avoid over-fitting, the loss on the validation dataset is computed four times per epoch and the model weights are stored as a checkpoint. Further, these validation results are used for early stopping of training when there is no improvement of the validation loss over three consecutive epochs. The best checkpoint is restored to form the final predictor.

## 5. Numerical Results

This section presents proof of concept results for the predictor model. We discuss the quality of trained models depending on the number of modes. Then for the selected seven-mode predictor, we show the convergence of training and its application on a test example. The example highlights the comparison to the deterministic predictor (single-mode MDN trained using the same procedure) on the noiseless and moderately noisy data.

### 5.1. Optimal Number of Modes

The optimal number of modes varies depending on the inverse problem and the dataset. In our case, the starting location and the well-path in the SVD coordinates are uncertain and multi-modal; these effects are further

**Table 1**
*Comparison of the Trained Models With a Different Number of Predicted Modes on the Test Dataset and With Manually Constructed SVD Functions*

| Dataset | Metric \ Modes | 1* | 2 | 3 | 4 | 5 | 6 | 7* | 8 | 9 | 0-base |
|---|---|---|---|---|---|---|---|---|---|---|---|
| Test dataset | NLL-mismatch | **0.3650** | 0.3817 | 0.4057 | 0.4193 | 0.4189 | 0.4355 | 0.4129 | 0.4185 | 0.4412 | 0.9444 |
| | NLL-mismatch (well-log) | 0.0067 | **0.0059** | 0.0068 | 0.0070 | 0.0064 | 0.0067 | 0.0063 | 0.0070 | 0.0071 | 0.0412 |
| | Best mode MAE | 2.34 | 1.60 | 1.32 | 1.25 | 1.21 | 1.24 | **1.13** | 1.13 | 1.20 | 6.04 |
| | Collapsed modes, % | 0.0 | 2.3 | 4.9 | 7.1 | 9.1 | 11.4 | 12.1 | 14.0 | 15.1 | – |
| Flat | Best mode MAE | 2.69 | 1.48 | 1.27 | 1.68 | 1.22 | **1.14** | 1.70 | 1.77 | 1.35 | 0.00 |
| Slope | Best mode MAE | 1.25 | 0.97 | 0.60 | 0.50 | 0.39 | 0.37 | **0.36** | 0.40 | 0.39 | 8.68 |
| Fault | Best mode MAE | 3.08 | 2.11 | 1.77 | 1.70 | 1.57 | 1.62 | 1.40 | **1.38** | 1.54 | 7.40 |
| Test dataset w. noise | NLL-mismatch | 0.8864 | 0.8041 | 0.8014 | **0.7473** | 0.8365 | 0.8335 | 0.7687 | 0.7749 | 0.8259 | 0.9444 |
| | NLL-mismatch (well-log) | 0.0206 | 0.0152 | 0.0170 | 0.0138 | 0.0157 | 0.0169 | **0.0135** | 0.0162 | 0.0177 | 0.0412 |
| | Best mode MAE | 5.67 | 3.65 | 3.04 | **2.66** | 3.02 | 3.01 | 2.75 | 2.73 | 3.01 | 6.04 |

*Note.* For the manually constructed functions: flat (constant 0); slope (straight line with inclinations of 82 and 98°); fault (same as slope, but with a single vertical fault with a through of 3.75 feet); the error is averaged over different offset well logs. The table also shows the results of an experiment with 1% noise added to the logs. The models marked with * are used further in Sections 5.3 and 5.4. 0-base shows a result for a constant baseline solution $b(x) = 0.0$. Best-mode MAE excludes the modes with less than 5% probability and is measured in computational cells. Bold numbers highlight the best metric in each of the rows.

increased to the prediction area not covered by data. A statistically-sound metric for multi-modal prediction is negative-log-likelihood (NLL) mismatch $I_{NLL}$, which is the classical loss function used for MDN:

$$I_{NLL} = -\log \sum_m p_m \exp\left( \frac{-\|b^* - b^m\|}{\sigma^*} \right),$$ (11)

where

$$\sigma^* = l^+ \alpha_{class} = 3.2$$ (12)

is mode-width normalization, the same as used in the training. This metric goes to zero when the prediction probability density converges to the correct solution. An MDN with standard random weights of the output layer results in $I_{NLL} \approx 10$, while a constant zero 'baseline' solution for the SVD function which dominates the dataset has $I_{NLL} \approx 0.9444$. As we can see from Table 1, regardless of the number of modes selected, the training results in reducing the NLL mismatch about 30 times compared to the random initialization and 2–3 times compared to the baseline solution. This is equivalent to either improving the error 2–3 times or improving the probability prediction exp(2) - exp(3) times. The NLL mismatch is slightly lower for the single-mode predictor as it is known to favor its average solutions (Cui et al., 2019). Thus we only use this statistically sound metric only for the evaluation and not for the training to avoid mode collapse observed by Earp and Curtis (2020); Meier et al. (2007).

It is often useful to capture several distinct modes in practical applications, including the interpretation of logs. To evaluate the realistic quality of prediction, we assess the predicted well-log instead using the same NLL mismatch metric, replacing $\|b^* - b^m\|$ by $\|g(b^*) - g(b^m)\|$ (only from 0 to $l$ where the data is sampled). The NLL mismatch for the predicted well-log is the smallest for the two-mode prediction, and 5, 6, and 7 modes also give a better well-log prediction than the deterministic version. This indicates that the deterministic model struggles to approximate the data but instead gives incorrect average solutions, which we further illustrate in examples in Subsection 5.3.

To assess the optimal number of modes needed for our problem we also evaluate all the models on the test data and three manually-defined scenarios of SVD functions: straight lines with inclinations of 90° (flat); 82 and 98° (slope); and straight lines with inclinations of 86 and 94, and a single vertical fault with a through of 3.75 feet. The prediction accuracy in all these scenarios (with different vertical reference well-logs) is also summarized in Table 1: the best mode distance excludes the modes with a probability of less than 5%. For the test dataset, the best-mode predictions tend to get better with the increase of the number of modes, with the best achieved for nine modes. We also see that the models with a smaller number of modes are not as good at resolving high-angle bedding and faulted scenarios. A total of 6-, 7-, and 8- modes are the best for the manually defined scenarios.

Finally, we count the number of 'collapsed' modes, which we define as pairs of modes that are closer to each other than the best mode is to the exact solution. The 7-mode model is the largest where the percentage of collapsed modes is lower than the average mode probability $1/7 \approx 14.3\%$, see Table 1. Considering this, and the observed good results in terms of other metrics, the seven-mode model is used in the rest of the results section as a reference model.

## 5.2. Convergence of the Predictor Model

The loss convergence for training and validation data of the seven-mode MDN are presented in Figure 3. The best-mode mismatch contribution (green curve) decreases through the whole training process, while the probability contribution (red curve) becomes stable after the first few epochs. The predictors with a different number of modes exhibit qualitatively similar training behavior. The entire training takes 5.2 hr on a single GeForce 2080ti GPU with PyTorch 1.9.0.

The quality of the probability prediction (classification contribution in the loss) can be assessed by counting how often a selected prediction is the closest to the actual solution on the validation dataset. A mode with a probability of, for example, 13% should correspond to the correct solution in approximately 13% of all cases. However, since our MDN predicts both the SVD functions and their probability, we cannot directly calculate such a metric. Thus, we resort to a law of large numbers to assess the probability prediction quality. We evaluate the model on the validation dataset and group predicted probability values into percentile buckets, that is, P0-P10, P10-P20,…and denote them, for example, $d_{10-20}$. We then count the number of best matches in each bucket; that is, when the best match is in, for example, P10-P20, we increment the counter denoted $a_{10-20}$. Assuming that the predicted probabilities are evenly distributed within each bucket, the fraction of the two numbers should represent the predicted percentile value $\tilde{P}$ and tend to the mean probability of the bucket, for example,:

$$\tilde{P}_{15} = \frac{a_{10-20}}{d_{10-20}} \rightarrow 15\%. \tag{13}$$

The convergence of bucketed probabilities on the evaluation dataset is shown in the bottom part of Figure 3. We see that the probabilities stabilize around correct values early during the training. Simultaneously the fraction of the matches in each bucket indicated by the colored area also stabilizes. We do not subdivide buckets above P30 due to the negligibly small number of samples and invalidity of the law of large numbers. The fact that they do not lie precisely in the center can be due to uneven probability prediction within each bucket.

We further quantify the accuracy of classification on the testing dataset. As before, we group the probabilities into percentile buckets. As shown in Figure 4, we plot the predicted percentile values against the actual percentile values. The perfect match will be obtained on the diagonal line going from zero. However, we expect some errors for the buckets with only a few samples. To visualize the severity of such a misfit, we draw error bars for each point with a length equal to $2/\sqrt{d}$, where $d$ is the number of samples in the bucket. In the lower part of the distribution, we again observe a good match. Starting from $\tilde{P}_{35}$, the predicted percentiles start deviating from the diagonal but remain within the error-bar limits, which are larger due to an insufficient amount of samples.

## 5.3. Application of the Model

In this subsection, we use a single SVD function b* from the testing dataset split into three consecutive examples and evaluate the predictor on successive pieces of that model. We note that the examples were selected to highlight the capability of a multi-modal predictor to capture several modes of the solution in unfavorable conditions. The average NLL mismatch for the presented examples is $I_{\text{NLL}} = 0.9418$, which is more than two times worse than the average recorded in Table 1.

Figure 5 shows the true SVD function b* and the predictions for segments separated by 32 ft from each other for ease of reading. The background shows the image composed of differences between the offset well-log and the horizontal well-log (4). The 0-iso-lines in the image correspond to well-log matches and thus are possible interpretations. However, the task of a predictor is to also account for the learned geological setting and output the seven most likely modes and their probabilities. All the predictions for each data point are generated by a single evaluation of the predictor, which takes about 1.5 ms.

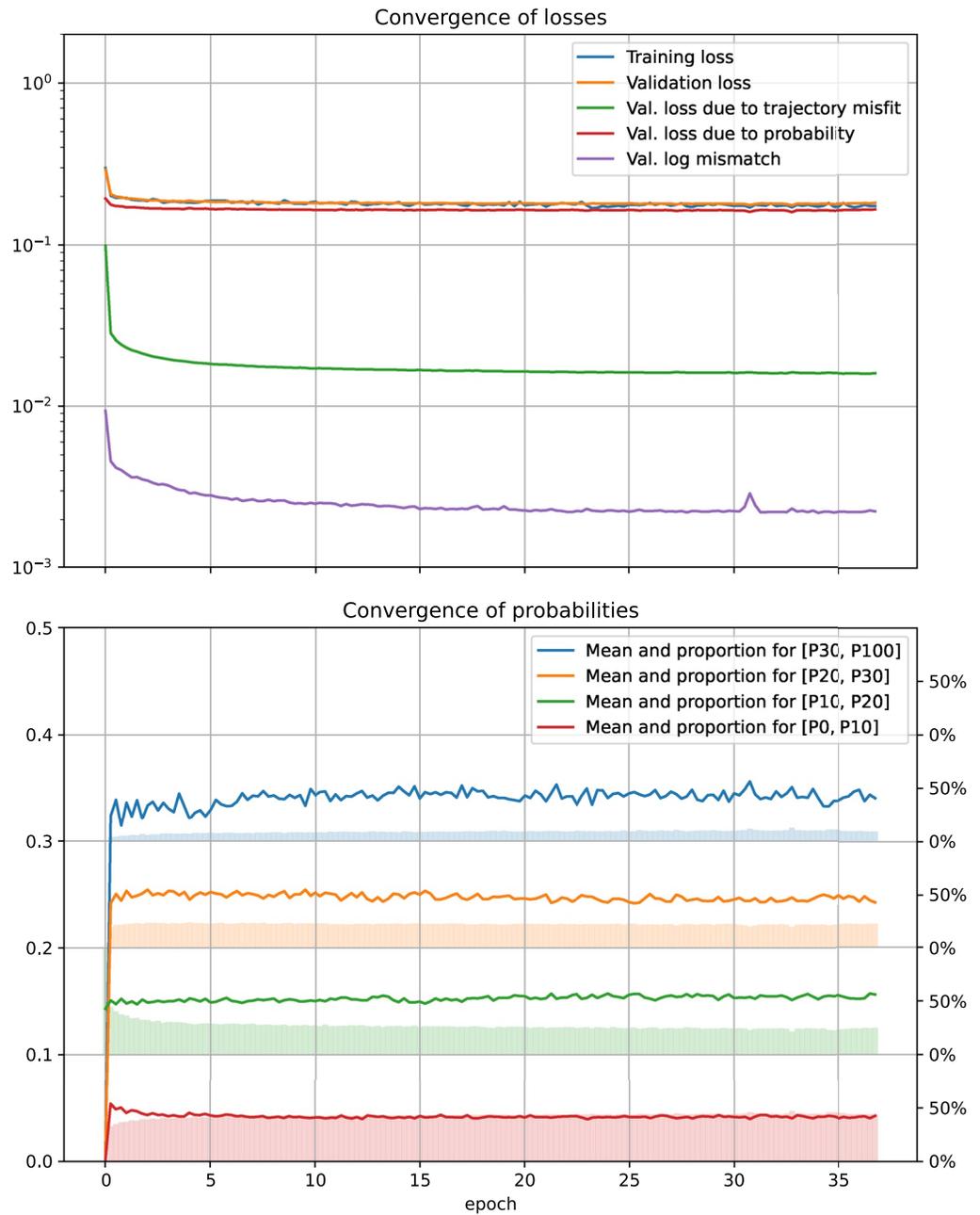

**Figure 3.** Top plot: The convergence of the losses and probabilities during training. The loss of the validation data is split into its regression (trajectory misfit) and classification (probability) contributions. The misfit of predicted well-log data is also shown. Bottom plot: The detailed convergence of the probabilities of the predicted modes on the validation dataset. All the probabilities ($7 * n_{samples}$) of the predicted modes are grouped into percentile buckets. The colored lines show the fraction of the modes in the bucket that give the closest match to the ground truth. The colored areas show the evolution of the fraction of samples in each bucket.

We see that the trained multi-mode model manages to predict the actual earth configuration with the accuracy of less than a cell for all three samples, even though the best prediction does not always have the highest probability. The single-mode predictor gets much worse results with an error of 2.0 cells or worse. Let us consider the individual samples which have complexity increasing from left to right:

The first sample is slightly inclined and centered around zero, thus representing the most common scenario in the training dataset (every second sample is shifted to zero). The multi-mode predictor identifies the two primary possibilities of bedding dipping up and down, however it predicts a higher probability of dipping up. The well-logs displayed in the bottom part of Figure 5 also show a good match for all modes inside the region with data. The curves overlay each other for the part with data, but the predictions diverge in the region with no data. The best matching mode resembles actual bedding accurately and has $p = 8\%$. Relatively low probability, in this case, can be explained by the good quality of the well-log data prediction of all the proposed solutions, each geologically realistic. The single-mode prediction also prioritizes dipping up, thus not capturing the correct mode of the data. Its predicted dip is not sufficiently steep toward the right.

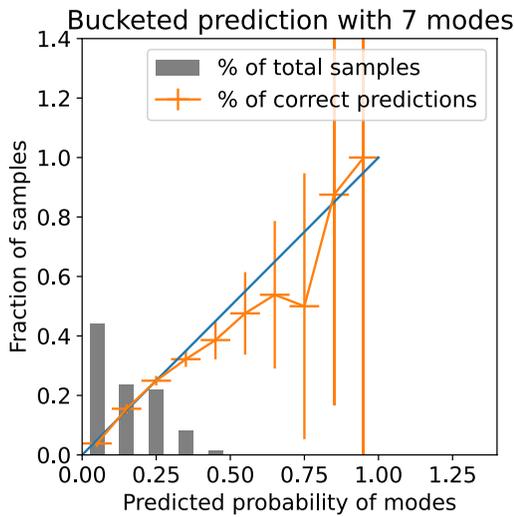

**Figure 4.** Accuracy of prediction of probabilities for the test dataset. The vertical error bars are proportional to the square root of samples in each bucket.

The second sample is curved bedding not centered around zero, which simulates a scenario where the previous inversion was not correct, which is slightly less common. Visual inspection of the input image shows four modes, two symmetric around 0.0 and two symmetric around 4.0. In this case, the best prediction has the second-highest probability of 17%. The highest probability prediction matches the well-log data and is very close to the single-mode prediction, which again picks the wrong mode in the input data. The closeness of single-mode and highest-probability multi-mode solutions indicate that the models capture similar trends embedded in the training data. The best prediction seems to neglect the first part of the well-log data below 40 ft but surprisingly manages to predict the bedding ahead. Less-probable modes b2, b5, and b6 are inferior in approximating the well-log data for this scenario. We also note that mode b7 starts where b2 left off in the first sample despite no information passed between the two evaluations. This demonstrates the robustness of prediction b2 to the area with no data.

The third sample contains a fault, which is less frequent in the dataset. Correct identification of the modes with different fault throws is specifically tricky. Nevertheless, the correct SVD function is resolved by b4, which also has the highest probability. The well-log-match quality in the presence of a fault is worse for all modes, but partial matches are observed for those with higher probability. The single-mode inversion seems to be close to the correct mode but again shows average behavior, which does not resemble the data very well.

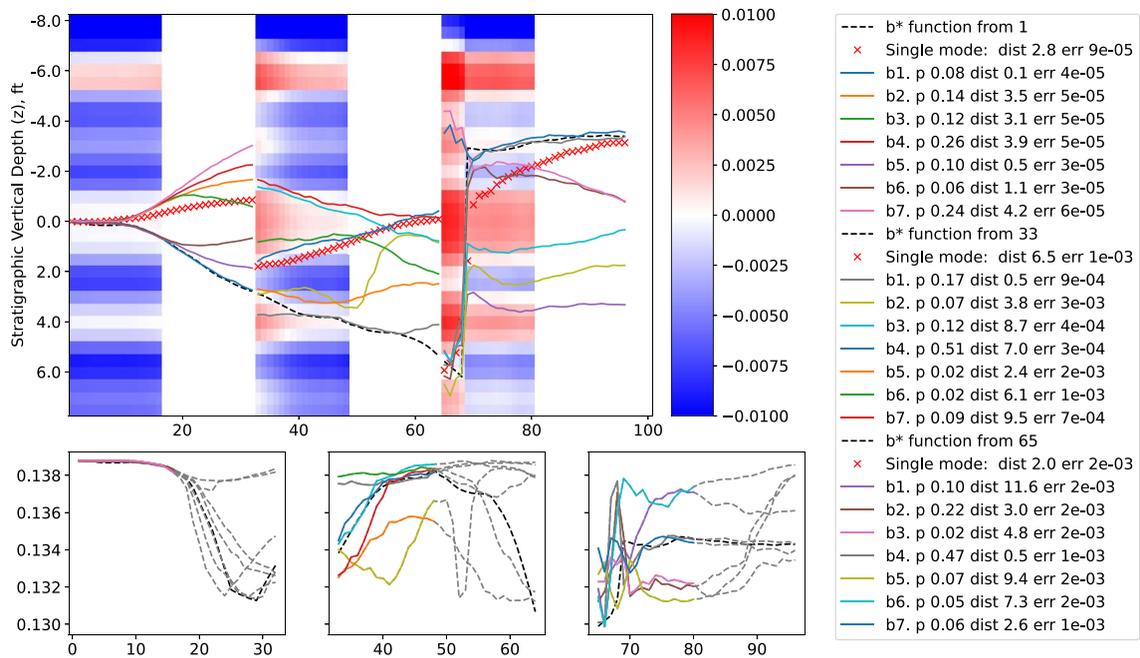

**Figure 5.** The prediction results for a multi-modal predictor on several consecutive samples from a test dataset. The true stratigraphic vertical depth function $b*$ is shown in dashed black. The background shows the input image. The seven inversions/predictions are shown in color, and their probabilities are in the legend. Deterministic (single-mode) predictions are shown for comparison. Abbreviations in the legend: the first index - node number, p - the probability of the mode, dist – mean-absolute-error (MAE) between the predicted curve and the truth (in cells), err - MAE between the well-log of the mode and the well-log of the truth. The second row shows the plots of well-logs for the first three well-log segments. The well-logs are extended to the prediction area in gray.

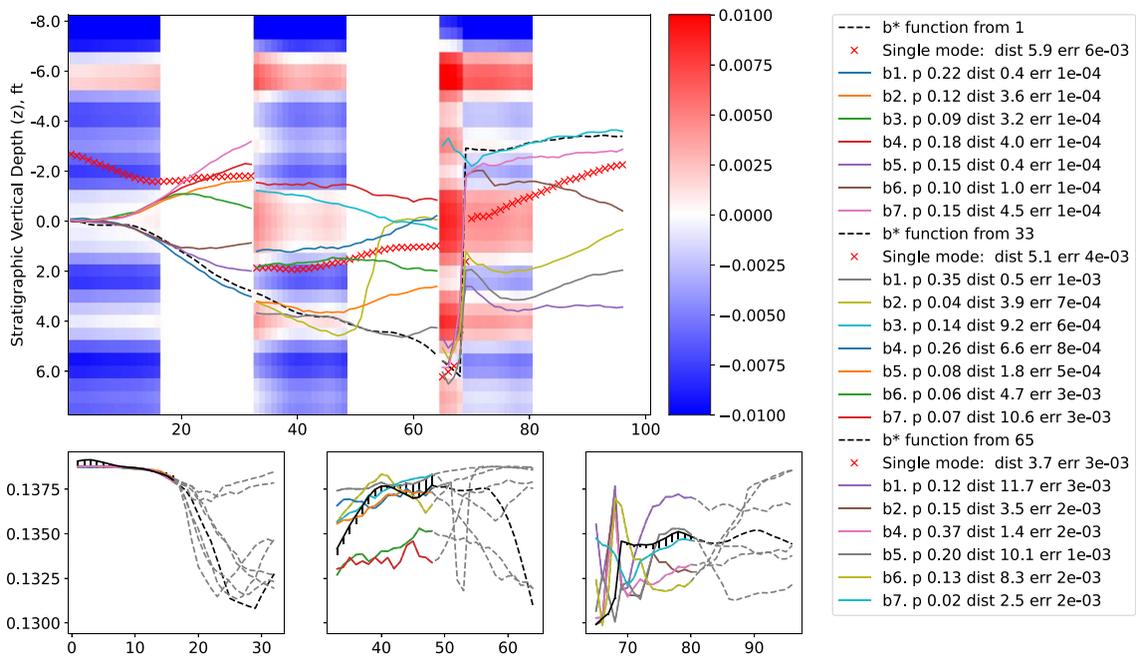

**Figure 6.** The inversion/prediction results on the data with 1% added noise using the same setup as in Figure 5. The second row shows the plots of well-logs for the first three well-log segments. The true noisy log is shown in black with error bars pointing to the corresponding well-log values before the added noise. The true and the predicted well-logs are extended to the prediction area as dashed lines.

### 5.4. Testing on Noisy Data

In this subsection, we test the multi-mode predictor on noisy data. We use the same test setup as in the previous subsection, but we add 1% correlated additive noise when generating the logs for the current well.

To simulate realistic noise, we first generate Gaussian white noise with a standard deviation equal to 1% of the range of the offset log. After that, we convolve the white noise with the exponential kernel $g$:

$$g(j) = \exp\left(\frac{-j^2}{2l_{corr}}\right), \tag{14}$$

which is computed for indices along $x$ in the range $[-l_{corr}, l_{corr}]$ with the correlation length $l_{corr}$ set to eight cells.

Figure 6 shows how the prediction result changes when the noise is added. Unlike the previous subsection (Figure 5), the true solution does not precisely follow the zero iso-lines in the heatmap image because of the added noise. Thus some of the inversions/predictions from the model differ significantly compared to the noiseless case. Nevertheless, for the first two samples without faults, the best prediction maintains the accuracy of one computational cell and is below two cells for the third case with fault. Moreover, the probability of the best-predicted mode is above 20% in all three samples, exceeding the metrics for the noiseless case. The single-mode inversion deteriorates significantly and provides worse results than in the noiseless case.

On the reconstructed logs, shown in the second row of Figure 6, it is interesting to observe that many predicted curves match the true offset log rather than the input log, which contains noise. The reason is the information about likely SVD function configurations embedded into the trained network. For the first segment, the multi-modal predictor inverts SVD functions starting from zero, rather than off-zero indicated by the data, which gives results similar to Figure 5. For the second segment, the prediction with the highest probability b1 deviates from the observed data but follows the true geology starting from about MD = 40. For the third example, the

**Table 2**
*Negative-Log-Likelihood Misfit of Seven-Mode Predictors on the Test Data With Different Noise Levels*

| 7-mode predictor | Evaluation noise | | | |
|---|---|---|---|---|
| Noise, training ↓ | 1% | 2% | 4% | 10% |
| *0-base-case* | 0.9444 | 0.9444 | 0.9444 | 0.9444 |
| *0% | 0.7687 | 1.0673 | 1.4507 | 1.8966 |
| 1% | 0.5609 | 0.6824 | 1.0129 | 1.5737 |
| 2% | 0.5592 | 0.6008 | 0.7497 | 1.2594 |
| 4% | 0.6226 | 0.6354 | 0.6809 | 0.9458 |
| 10% | 0.7290 | 0.7322 | 0.7429 | 0.7996 |

*Note.* Additionally to the reference Model for this Study Trained With 0% Noise, the Table Includes MDNs Trained on Noisy Data. 0-Base-case Shows a Result for a Constant Baseline Solution $b(x) = 0.0$. The Diagonal Dotted Line Indicates the Cases Where the Same Noise Level was Used During Training and Testing.

correct fault mode seems to be recovered approximately despite quite a significant data mismatch in the predicted logs.

The observed improvements in the probability predictions over the noiseless case for the multi-mode predictor seem counter-intuitive since the MDN was not trained on noisy data. However, this is a coincidence for the selected example which is challenging for the model. In the following subsection, we further analyze the impact of increasing noise on the training and testing of the model.

### 5.5. Impact of Noise Level

The average scores of the reference model on the 1%-noise test described above are about two times worse in the considered metrics (see Table 1). In an attempt to improve the model's performance, we introduce noise during the training of seven-mode MDN in this subsection and consider its impact on the model's performance on noisy data. We use the correlated noise obtained by convolution with the exponential kernel described by Equation 14 for training and evaluation.

Table 2 shows cross-testing of seven-mode MDN models trained and evaluated with different noise levels from 0% to 10%. The performance *0%-noise-model, used as the reference in the rest of the study, goes below acceptable (defined by the uninformed base case) when testing with noise above 2%.

Adding the noise during training helps the MDN to overcome this limitation and deliver acceptable performance on noise levels above 2% (see Table 2). The noise level increase during training shifts the priority from learning to match the hard data, aka logs, to learning the geological configurations. Unsurprisingly, the model trained with the corresponding noise level exhibits the best performance for a given noise level in each column. The results of the models trained with slightly higher noise levels are only marginally worse. The 2%-noise-model even performed somewhat better on 1%-noise testing. This improvement might be related to the positive effect of added noise improving stochastic gradient descent observed in the study by Neelakantan et al. (2015). On the contrary, reducing the noise level during training by roughly 3% compared to the testing noise gives the NLL mismatch worse than the 0-base-case model.

Following the rows given in Table 2, we see that the models' performance deteriorates to the right from the diagonal (when evaluated with higher noise levels than used in training). A noise increase of roughly 3% during testing renders performance unacceptable. On the other hand, the NLL mismatch decreases when reducing the noise compared to the level used during training. Since all models learn the log matching ability, they take advantage of better quality, less noisy data to give better predictions.

The best-case performance decreases with each row as the models gradually reduce the 'trust' in hard data. Nevertheless, the model trained with a noise level of 10% across the tested noise levels is on-par with the good results of the noise-less model challenged by the 1% noise described in the previous subsection. Thus the proposed method can handle the log noise above the industry standard of 6% to 7% used for the Geosteering World Cup (Tadjer et al., 2021).

## 6. Conclusions

In this work, we have presented a new deep-learning methodology for direct multi-modal inversion of geophysical logs. The training of neural-network-based predictors is performed offline. A single evaluation of the trained network generates a predefined number of SVD functions, containing both interpretations and predictions of stratigraphy ahead of data. The evaluation takes 1.5 ms compared to the "minutes" reported for the algorithm from Maus et al. (2020) for multi-solution tracking, which is already orders of magnitude faster than the manual workflows taking hours.

During the initial testing of the model on gamma-ray well-logs, we achieved an average best-case inversion/prediction with an MAE of less than 1.5 computational cells. This error includes half of the trajectory, extrapolated ahead of data. We observe that non-optimal predictions with high probability also give a good match of well-log data, thus resolving plausible modes. Key to this performance is the ability of the MDN to learn both: to interpret the well-log data, and predict geology based on the frequency of different geological scenarios in the training dataset.

When trained on noisy data, the presented methodology can handle realistic log-noise levels. Thus the model's sequential version (Alyaev et al., 2022) can be directly used for automatic interpretation during traditional shallow-log-based geosteering with fixed-thickness stratigraphy and possible faults. We expect the approach to naturally extend to other geological discontinuities such as pinch-outs if retrained on a representative dataset. Furthermore, the presented methodology can have implications for other multi-modal problems in geophysics and geosciences for which other Bayesian methods are currently used. They include reservoir-mapping inversion of deep electromagnetic wellbore measurements (Shahriari et al., 2021), seismic waveform inversion (Mosser et al., 2020), magnetotelluric data inversion (Xiang et al., 2018), and permeability estimation (Wang et al., 2021).

## Data Availability Statement

The data used in this study is available from Alyaev (2022) and Miner et al. (2021). For reproducibility, our implementation of the MTP loss function is openly available (Alyaev & Elsheikh, 2022).

## References

**Acknowledgments**

This work is part of the Center for Research-based Innovation DigiWells: Digital Well Center for Value Creation, Competitiveness and Minimum Environmental Footprint (NFR SFI project no. 309589, DigiWells.no). The center is a cooperation of NORCE Norwegian Research Centre, the University of Stavanger, the Norwegian University of Science and Technology (NTNU), and the University of Bergen. It is funded by Aker BP, ConocoPhillips, Equinor, Lundin Energy, TotalEnergies, Vår Energi, Wintershall Dea, and the Research Council of Norway.

We would like to thank four anonymous reviewers for their suggestions and comments on earlier versions of this study.